\documentclass[pra,aps,twocolumn,superscriptaddress,showpacs,reprint]{revtex4}
\usepackage{amsfonts}
\usepackage{amsmath}
\usepackage{amssymb}
\usepackage{graphicx}
\usepackage{epstopdf}
\usepackage{subfigure}

\providecommand{\U}[1]{\protect\rule{.1in}{.1in}}

\begin{document}

\preprint{HEP/123-qed}
\title[ ]{ The origin of the Redshift Spikes in the reflection spectrum of a
Few-cycle Pulse in a Dense Medium}
\author{Yue-Yue Chen}
\affiliation{State Key Laboratory of High Field Laser Physics, Shanghai Institution of
Optics and Fine Mechanics, Chinese Academy of Sciences, Shanghai 201800,
China}
\author{Xun-Li Feng}

\affiliation{Department of Physics, Shanghai Normal University, Shanghai 200234, China}

\author{Zhi-Zhan Xu}

\affiliation{State Key Laboratory of High Field Laser Physics, Shanghai Institution of
Optics and Fine Mechanics, Chinese Academy of Sciences, Shanghai 201800,
China}
\author{Chengpu Liu}
\email{chpliu@siom.ac.cn}
\affiliation{State Key Laboratory of High Field Laser Physics, Shanghai Institution of
Optics and Fine Mechanics, Chinese Academy of Sciences, Shanghai 201800,
China}

\begin{abstract}
We give a detailed description about the reflected spectrum of a few-cycle
pulse propagating through a resonant dense medium. An unexpected
low-frequency spike appeared in the red edge of the spectrum. To figure out
the origin of this redshift spike, we analysis the mechanisms responsible
for the redshift of the reflected field. So far, the redshift has not been
well studied for few-cycle pulses except a brief explanation made by the
previous study [Kaloshan $et$ $al$., Phys. Rev. Lett. 83 544 (1999).], which
attributed the origin of the redshift to the so-called intrapulse four-wave
mixing. However, we demonstrate numerically that the redshift consists of two
separated spikes is actually produced by the Doppler effect of
backpropagation waves, which is an analogue effect of dynamic nonlinear
optical skin effect. Our study elucidates the underlying physics of the
dynamic nonlinear optical effects responsible for the redshift spikes.
Moreover, the dependency of the their frequency on the laser and medium parameters, such as medium density
and input pulse area are also discussed.
\end{abstract}

\volumeyear{year}
\volumenumber{number}
\issuenumber{number}
\eid{identifier}
\date[Date text]{date}
\received[Received text]{date}
\revised[Revised text]{date}
\accepted[Accepted text]{date}
\published[Published text]{date}
\startpage{1}
\endpage{102}
\maketitle

With the rapid development of ultrafast science technology \cite{Zhou,Nisoli}%
, the field of optics soon entered the new era of extreme nonlinear optics.
The interaction between a few cycles pulse with less than 5fs duration and
resonant two-level system gives rise to a wealth variety of new phenomena,
effects and applications. In this regime, the standard approximations used
in the traditional nonlinear optics are no longer appropriate \cite%
{Rothenberg}. In this case, the full Maxwell-Bloch (MB) equations without
slowly varying envelop approximation (SVEA) and rotating wave approximation
(RWA) need to be solved, which can be done by an iterative
predictor-corrector finite-difference time-domain method. Backpropagation is
result from the self-reflection occurred in a saturable medium. The ability
to solve the full Maxwell equation numerically fuels the study on the
backpropagation of pulse ignored in the ``one-way going" approximation \cite%
{Bullough} which is implied in envelope forms of MB equation. However, for a
dense medium, backpropagation may has a significant impact on both the
reflected and transmitted field, and a vast variety of intriguing new
physical phenomena are emerged such as dynamic nonlinear optical skin (DNOS)
effect \cite{Forysiak} and decay of self-induced-transparency pulses \cite%
{Robert}.

The reflection spectrum of the few-cycle pulses has many unique features,
and the most conspicuous one is the redshift observed in \cite{Kalosha}. We
also notice that, at the redside of the spectrum, an unexpected
low-frequency spike is located at the red edge, along with another peak with
higher frequency. To tell the mechanism behind those redspikes, we resort to
the existing few-cycle theories about the physical interpretation of the
observed redshift given by \cite{Kalosha,Xie}, where the redshift is briefly
attribute to intrapulse third-order four-wave mixing(FWM). However, the
obtained scaling laws in terms of laser or medium parameters are
inconsistent with FWM theory in so many ways. On the contrary, both the
reflected field profile and spectrum can be explained very well with the
Doppler effect of the backpropagation. This theory is based on the fact that
the self-reflect interface is propagating through the medium with the
transmitted pulse, which acts as a moving mirror inducing a redshift in the
backpropagation waves \cite{Forysiak}.

In this letter, we fully study the mechanism of the redshift observed in the
reflection spectrum of a few-cycle pulse propagating through a dense
two-level atomic medium (DTLA). Our study is focused on the redshift spikes
observed in the reflection spectrum and make clear the underlying physics by
a theory based on the doppler effect of moving self-reflection interface
instead of FWM. For the redspike with the largest shift, we also show how to
change its location and amplitude by varying laser or medium parameters.

\begin{figure}[tbp]
\begin{center}
\includegraphics[width=
0.5\textwidth]{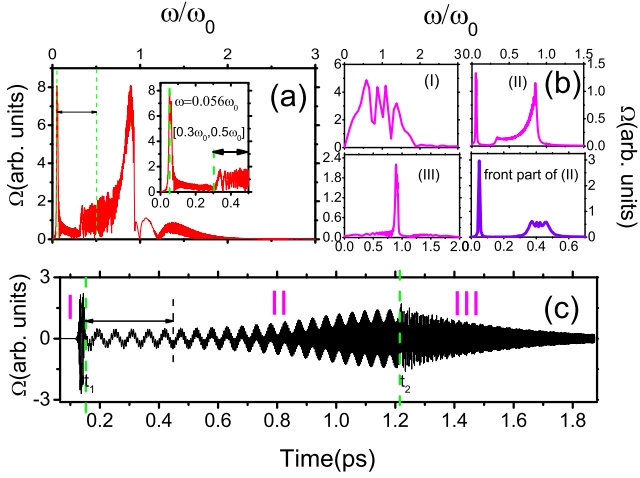}
\end{center}
\caption{ Electric field profile and spectrum for three successional time
regions. $\protect\omega _{c}=0.1$fs$^{-1},A=4\protect\pi ,L=150\protect\mu $%
m$,t_{1}=0.15$ps$,t_{2}=1.21$ps$.$ $\left[ t_{i},t_{1}\right] ,\left[
t_{1},t_{2}\right] $ and $\left[ t_{2},t_{f}\right] $ are labeled by I,II
and III respectively. The below insets are the corresponding spectrums for
those three regions, while the above inset is the spectrum of the first five
oscillations in region II.}
\end{figure}

Assume the electromagnetic field is linearly polarized $%
E=E_{x}(z),H=H_{y}(z).$ The spatial orientation of the dipole is along $x,$
and the macroscopic nonlinear polarization $P=P_{x}(z).$ The Maxwell
equations for the DTLA medium takes the form \cite{Allen}
\begin{eqnarray}
\partial _{t}H_{y} &=&-\frac{1}{\mu _{0}}\partial _{z}E_{x}, \\
\partial _{t}E_{x} &=&-\frac{1}{\epsilon _{0}}\partial _{z}H_{y}-\frac{1}{%
\epsilon _{0}}\partial _{t}P_{x}.  \notag
\end{eqnarray}%
The off-diagonal density matrix element $\rho _{12}=(u+i\upsilon )/2,$ and
the population inversion $w=\rho _{22}-\rho _{11}$ between excited and
ground state. $u,\upsilon $ and $w$ obey the following set of Bloch
equations,
\begin{eqnarray}
\partial _{t}u &=&-\gamma _{2}u-\omega _{0}\upsilon , \\
\partial _{t}\upsilon &=&-\gamma _{2}u+\omega _{0}\upsilon +2\Omega w,
\notag \\
\partial _{t}w &=&-\gamma _{1}(w-w_{0})-2\Omega \upsilon .  \notag
\end{eqnarray}%
Where $\gamma _{1},\gamma _{2}$ are, respectively, the population and
polarization relaxation constants, $\omega _{0}$ is the resonant frequency, $%
\Omega (\Omega =dE_{x}/\hbar) $is Rabi frequency, and $w_{0}$ is the initial
population difference$.$ The MB equation can be solved by adopting Yee's
finite-difference time-domain discretization method for the electromagnetic
fields and the predictor-corrector method for the medium variables \cite%
{Ziolkowski,Hughes,Yee,Taflove}. The initial condition for the input pulse
is $\Omega (t=0,z)=\Omega _{0}\cos [\omega _{p}(z-z_{0})/c]\sec
h[1.76(z-z_{0})/(c\tau _{p})],$ where $\Omega _{0}$ is the peak Rabi
frequency for the input pulse, $\tau _{p}$ is the full width at half maximum
(FWHM) of the pulse intensity envelop, and the initial position $z_{0}$ is
set to be a value large enough to avoid the pulse penetrating into the
medium at $t=0$. The medium is initialized with $u=\upsilon =0,w_{0}=-1.$ To
study the reflection of the pulse, we adopt the following parameters to
integrate MB equation: $\omega _{0}=\omega _{p}=2.3$fs$^{-1},$ $%
z_{in}=52.5\mu $m$,z_{out}=202.5\mu $m$,z_{0}=26.25\mu $m, $d=2\times 10^{29}
$Asm, $\gamma _{1}^{-1}=1$ps$,\gamma _{2}^{-1}=0.5$ps$,\tau _{p}=5$fs$%
,\Omega _{0}=1.408$fs$^{-1},$the corresponding pulse area $A(z)=d/\hbar
\int_{-\infty }^{\infty }E_{0}(z,t^{^{\prime }})dt^{^{\prime }}=\Omega
_{0}\tau _{p}\pi /1.76=4\pi .$ Define a collective frequency parameter $%
\omega _{c}=Nd^{2}/\epsilon _{0}\hbar =0.1$fs$^{-1}$ to represent the
coupling strength between medium and field.

With the above parameters, we obtain the reflection spectrum as shown in
Fig.1(a). It can be seen, the spectrum is generally composed of two red
peaks, a sharp low -frequency spike and a broader one with higher frequency.
Here we denote this two spikes as redspike I and II. The unexpected redspike
I appears in the red edge centered at $\omega=0.056\omega_{0}$ with a FWHM $%
\tau=0.01\omega_{0}$ as shown in the inset of Fig.1(a). While the other has
a wider frequency ranging from $0.3\omega_{0}$ to $\omega_{0}$. To figure
out the origin of these redspikes, we divide the reflected pulse into three
successional time regions according to the electric profiles, as shown in
Fig.1 (c). The first three pictures in Fig.1 (b) show the corresponding
spectrums of the three regions and last one is the spectrum of the fist five
waves in region II. It can be seen both redspike I and II appear in spectrum II,
which are corresponding to a low-frequency modulation and a time-dependent frequency generated during the propagation in region II, respectively.
In the following text, we
analysis the underlying physics behind the electric profile in each region
to reveal the origin of these redspikes.

\begin{figure}[htbp]
\begin{center}
\includegraphics[width=
0.5\textwidth]{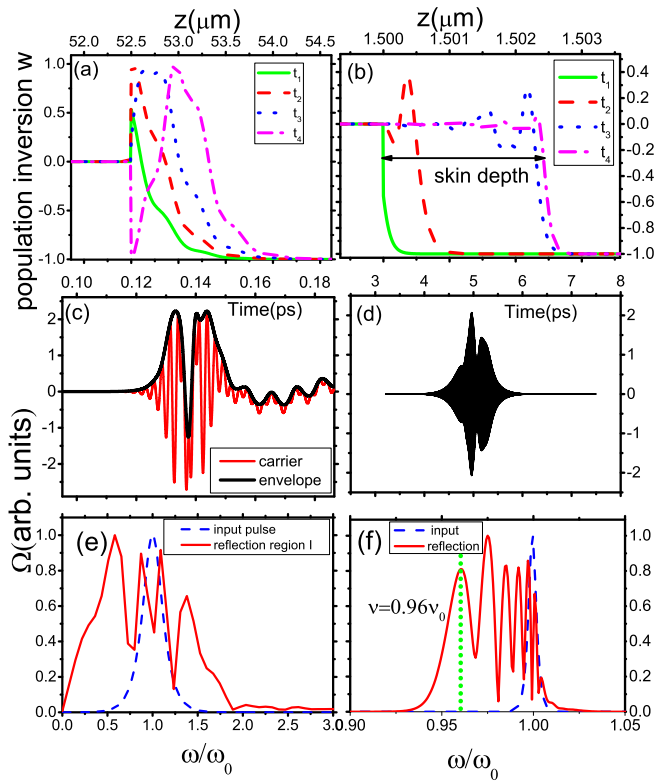}
\end{center}
\caption{(color online)  The compairation between a few-cycle pulse in
region I (a)(c)(e) and a long pulse (b)(d)(f) in terms of population
difference (a)(b), reflected field profile (c)(d) and spectrum(e)(f). The
parameters for the few-cycle pulse  $\protect\tau _{p}=5$fs$,A=4\protect\pi ,%
\protect\omega _{c}=0.1$fs$^{-1},\protect\lambda _{0}=830$nm$,$ $t_{1}=83.75$%
fs (solid line)$,t_{2}=85$fs(dashed line)$,t_{3}=86.25$fs(dotted line)$%
,t_{4}=87.625$fs(dashed-dotted line)$;$ a long pulse  $\protect\tau _{p}=300$%
fs$,A\approx 10\protect\pi ,\protect\omega _{c}=6.86$fs$^{-1},\protect%
\lambda _{0}=942$nm$,t_{1}=2.16$ps$,t_{2}=2.415$ps$,t_{3}=2.755$ps,$%
t_{4}=3.095$ps. The skin depth is the location corresponding to $t_{4}$. The maximum
velocity for (a) and (b) is $\upsilon_{m}=0.68c$  and $\upsilon_{m}^{l}=0.0183c$ respectively.
The corresponding redshift is $\nu_{m}=0.3\nu_{0}$ and $\nu_{m}^{l}=0.96\nu_{0},$ respectively}
\end{figure}

In region I, when a few-cycle pulse propagates along +z in vacuum to a
surface of a spatially homogeneous DTLA materials, an absorption front is
built near the surface, which separates the excited and ground state, as
shown in Fig.2(a). If the induced nonlinear polarization is sufficiently
large, which can be achieved by raising density, to result in spatial
inhomogeneities $n(\omega )\rightarrow n(\omega ,z)$ rapidly, a wave
propagating along -z may arise because of the constructive interference of
reflections from individual spatial inhomogeneities \cite{Robert}. Fig.2 (c)
depicts the profile of the reflected field in region I, a ringing appears
owing to the interference between the incident and the backpropagation
waves. From the corresponding spectrum illustrated in Fig.2 (e), we can see
that a remarkable redshift is observed in the spectrum with considerable
spectral broadening and modulation compared with the input spectrum. The
characteristics of the profiles and spectrum remind us of DNOS effect used
to explain a slight redshift in the reflection spectrum of long pulse, as
shown in Fig.2(f). Due to the similarity between this two cases as shows in
Fig.2, we extend the DNOS theory to the regime of extreme nonlinearity. That
is, the redshift produced in region I can be explained by the motion of the
absorption front near surface instead of FWM.

Several unique features of the few-cycle cases have to be emphasized. Since
the few-cycle pulse is more robust against reflection due to the high
intensity, it can propagate through the medium with a higher speed instead
of penetrating within a nonlinear skin depth as shown in Fig.1 (a) and (b).
The velocity of the moving
front is close to $c$, which invalids the image-source method used to derive
the equation $\Delta \lambda /\lambda _{0}=2\upsilon /c$ used in DNOS \cite%
{Sommerfeld}$.$ Thus, we use a substituted equation $\nu =\frac{c-\upsilon }{%
c+\upsilon }\nu _{0}$ to obtain the redshift of the reflected pulse due to
the Doppler effect for $\upsilon$ is close to c \cite{Doppler}. In region I, sample
two random times, we get $\upsilon _{1}=2.026\times 10^{8}$m/s, $\upsilon
_{2}=1.61\times 10^{8}$m/s, the corresponding frequencies in reflection
spectrum are $\nu _{1}=0.3\nu _{0},\nu _{2}=0.53\nu _{0}.$ Compared
Fig.1 (a) and (b) spectrum II, it can be inferred, in general, the function of region I
is adding local oscillations to the original smooth spectrum produced in
region II. Moreover, owing to the ultrashort and intense features of the
few-cycle pulses, effects such as the carrier Rabi flopping, self-phase
modulation and intrapulse FWM may take place, which enrich the spectrum of
region I in terms of the appearance of blueshift as shown in Fig.2 (e) and (f).

Now let us move on to the second region, where DNOS effect itself is not
sufficient to explain the electric profile. In this region, multiple Rabi
flopping causes 4$\pi $ pulse to split into two 2$\pi $ pulse [16]. The
first one is more intense and propagates much faster than the second one.
These two splitted 2$\pi $ pulses completely excite and deexcite the DTLA
medium in different locations, resulting in two moving absorption fronts
which travel with the 2$\pi $ pulses respectively. The frequencies of the backpropagation waves
are obviously velocity-dependent. Those two low-frequency backpropagation
waves may interfere with each other, giving rise to the electric profile
shows in region II. Note that, the remarkable difference between the region
I and region II is the number of effective self-reflected interface and
thus the number of the low-frequency components. Specifically, just one interface appears
in the former while for the latter it is
relied on the area of the input pulse. Also, the first 2$\pi $ pulse propagates much faster than
the 4$\pi $ pulse penetrating through the interface, such that the
self-reflected wave related to the first 2$\pi $ pulse possess a much larger
redshift than that in region I.

At the beginning of the region II around $0.15$ps, the absorption fronts are
moving forward with a rate $\upsilon _{1}=2.66\times 10^{8}$m/s and $%
\upsilon _{2}=1.33\times 10^{8}$m/s for the first and second pulses,
respectively. With the propagation proceeding, the velocity changes of the
first 2$\pi $\ pulse is negligible while the second one experiences a
deceleration with $a\approx -1.74\times 10^{20}$m/s$^{2}$. For the first
pulse, $t=0.15$ps, $\upsilon _{1}=2.66\times 10^{8}$m/s, $\nu
_{1}=0.06v_{0};t^{^{\prime }}=0.643$ps, $\upsilon _{1}^{^{\prime
}}=2.74\times 10^{8}$m/s, $\nu _{1}^{^{\prime }}=0.045\nu _{0},$ where $t$
and $t^{^{\prime }}$ present the moment region II begins and that
one of the pulse is outside of the medium, respectively. Note that,
redspike I is right centered at $\nu =0.056\nu _{0}$ with $\tau=0.01\nu _{0}$
as mentioned above. This indicates that the motion of
absorbing front induced by the first 2$\pi $ pulse is responsible for the
emergence of the redspike I. By that analogy, for
the second 2$\pi $ pulse $t=0.15$ps, $\upsilon _{1}=1.33\times 10^{8}$m/s, $%
\nu _{1}=0.39v_{0};t^{^{\prime }}=0.643$ps, $\upsilon _{1}^{^{\prime
}}=0.3\times 10^{8}$m/s, $\nu _{1}^{^{\prime }}=0.8v_{0},$ it's fair to
extrapolate that redspike II is related to the
propagation of the second 2$\pi $ pulse and its speed variation is the root
of the spectrum broading of this spike.
\begin{figure}[bp]
\begin{center}
\includegraphics[width=
0.5\textwidth]{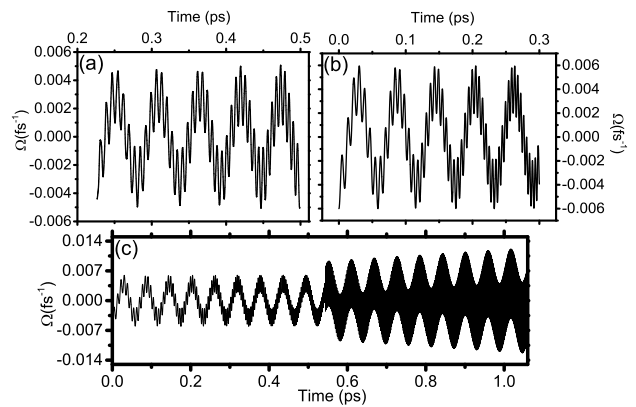}
\end{center}
\caption{The comparison of reflected field profiles between numerical result
and interference model. (a) is the electric profile during the first several
ps in region II, (b) is the counterpart obtained by interference model.
(c)is the longer time version of the interference model.}
\end{figure}

To confirm our theory, we give a interference model for the backpropation
waves. $\Omega (z,t)=\Omega_{01}\cos (\omega _{1}t+k_{1}z)\hat{e}_{z}+\Omega
_{02}(t)\cos (\omega _{2}(t)t+k_{2}z)\hat{e}_{z},$where $\upsilon
_{2}(t)=\upsilon _{20}-at,\omega _{2}(t)=\frac{c-\upsilon _{2}(t)}{%
c+\upsilon _{2}(t)}\omega _{0},\omega _{1\text{ }}$and $\omega _{2}(t)$ $%
(\Omega _{01}$ and $\Omega _{02}(t))$ are the frequencies (amplitude profiles) of the
reflected electric field of first and second $2\pi $ pulses, respectively. $%
\upsilon _{1\text{ }}$and $\upsilon _{2}(t)$ are the speed of the absorbing
front propagating along $+z$, $z=0$ is where detector is placed. Substitute
the variables with the following parameters: $\upsilon _{20}=1.3\times 10^{8}
$m/s, $a=-1.74\times 10^{20}$m/s$^{2},\omega _{0}=2.3$fs$^{-1},\Omega _{01}$%
=4$\times 10^{12}$Hz. $\Omega _{02}=$2$\times 10^{12}$Hz before the first
pulse propagates outside the medium and begins to increase due to the
influence of the reflection of the first pulse occurred in the back
interface. Thus, we obtain the electric profile in Fig.3. Our interference
theory is qualitatively in accordance with the numerical results.

In region III, the reflected fields of the first pulse by output interface
approach the detector and destroy the profile in region II. Initially,
redshift is disappeared in the spectrum due to the immobilized reflector,
then the fields are suffered from absorption during the propagation to the
detector, leaving a hole in the spectrum at $\omega =\omega _{0}.$ When this
reflected fields meet with the back waves produced by the second pulse, the
redshift with relatively high frequency is visible in the spectrum, as shown
in Fig.1 (b) III.

To sum up, the redshifts are result from the Doppler effect induced by
the motion of the self-reflection interface. The previous suggestions that both
the redshifts and bluehifts in reflected and
transmitted fields are the results of intrapulse FWM is questionable, if not
inaccurate. To be more persuasive, we discuss the factors that affect the
redshifts and predict the changes of the redshift in spike I with laser and
medium parameters

First, we discuss the impact of density by reducing the density.
Fig. 4(a) and (b) depict the variation of the transmitted and reflected
spectrum with the density is  decreased
from $\omega _{c}=$1.0fs$^{-1}$ to $\omega _{c}=$0.05fs$^{-1}$.
Fig. 4 (c) illustrates the dependence of redshift of spike I on density, which
show decrease in a linear pattern.
It can be seen, the redspike I in Fig.4 (b) is moving towards the redside with a declining amplitude,
while the bulueshift is tend to be diminished as shown in Fig.4 (a).
Interestingly, the responses of the redshift and blueshift to the density
reduction is quite opposite, which suggests that the physics behind them may not be
uniformed. Abandon the idea that both the shift is result from FWM,
the observed change rules can be explained very well with the velocity-dependent theory.
Specifically, since the red spikes are produced by the propagation of the
splitted pulses, lowing density will certainly increase their speed. The
motions of the complete population inversion induced by the pulses also
speed up. Therefore, the redshift is increased in the consequence of the
acceleration of the reflector.
\begin{figure}[tbp]
\begin{center}
\includegraphics[width=
0.5\textwidth]{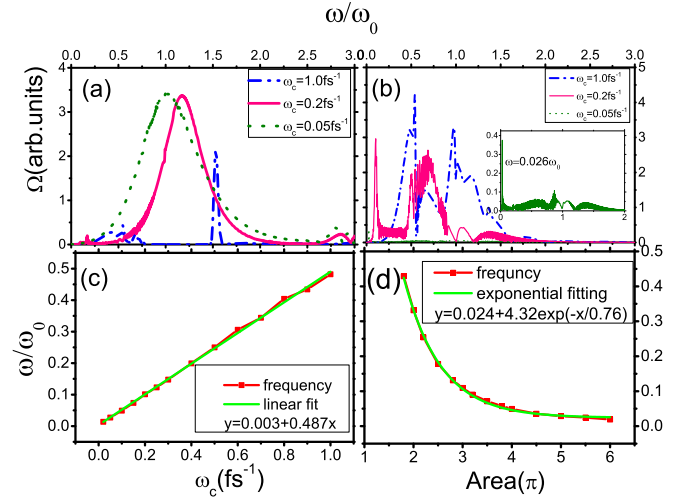}
\end{center}
\caption{(color online) The transmitted (a) and reflected (b) spectrums for
different density, $\protect\omega _{c}=$1fs$^{-1}$(dotted-dashed line), $%
\protect\omega _{c}=$0.2fs$^{-1}$(solid line), $\protect\omega _{c}=$0.05fs$%
^{-1}$(dotted line), with L=150$\protect\mu $m$,\Omega =1.4$fs$^{-1}$. The
inset in (b) is the enlarged view of $\protect\omega _{c}=$0.05fs$^{-1}.$
(a)-(c) are the changes of amplitude (solid line) and frequency (dashed line)
of the redspike I with medium density and pulse area. The other
parameters are L=45$\protect\mu $m$,\Omega =1.4$fs$^{-1},$ $\protect\omega %
_{c}=$0.2fs$^{-1}$, $\Omega =1.4$fs$^{-1}$.}
\end{figure}

Fig.4(d) reveals the scaling law for laser area. With the increasing of the input
pulse area, the frequency of the redspike I experiences a exponential decline.
This is because, with a fixed duration $\tau_{p}$, increasing the area is actually
increasing the intensity. The intensity and duration of the first 2$\pi$ pulse
obtained from Rabi flops is dependent on the intensity of the input pulse.
The more intense the input pulse is, the shorter and more intense the first 2$\pi$ pulse is.
As is known, the shorter intense 2$\pi$ pulse can exchange energy with medium rapidly
and thus propagate with a higher speed, which is responsible
for a lager redshift.

In conclusion, we give a thorough discussion about the reflected fields of
a few-cycle pulse propagation through a DTLA medium and the physics
mechanism behind the whole propagation process. The Doppler effect induced
by the moving reflector both near and inside the medium are responsible for
the redshift in the spectrum. Few-cycle pulses in region I can also have a DNOS alike
effect, which results in the interference between input and reflected fields.
The reshift produced during this region is relatively small due to the lower
speed compared with that in region II. The interference of the two backward
propagation waves produces the electric profile observed in the region II.
For a $4\pi$ pulse, the redshift spike with the largest shift is related to the first 2$\pi $
pulses, while the second redshift with higher frequencies is related to the
second pulse.
With the knowledge that the origin of the redspikes appear in the reflected spectrum
is related to the doppler shift of backpropagation waves,
the redshift can be controlled in terms of its location and amplitude by varying
parameters such as medium density and laser intensity to change the reflector. Material structures that are specifically
designed to broken the self-reflect interface can also be used to suppress
redshift.


\begin{thebibliography}{99}
\bibitem{Doppler} {\ The mirror moves away from the source with velocity $%
\upsilon ,$ and wave velocity is $u,u>\upsilon .$ Assume the first wave
front is reflected at time $t$, and the second wave front meets the mirror
at $t+\tau .$ In this case, $\tau =\frac{\lambda _{0}}{u-\upsilon }$ and the
wavelength $\lambda $ of the reflected light by a moving mirror\ is the
distance between second and first wave front at $t+\tau ,\lambda =\frac{%
u+\upsilon }{u-\upsilon }\lambda _{0},\nu =\frac{u}{\lambda }=\frac{%
u+\upsilon }{u-\upsilon }\nu _{0}.$}
\end{thebibliography}
\end{document}